\documentclass[3p,times]{elsarticle}

\usepackage{ecrc}
\usepackage{epstopdf}

\volume{00}

\firstpage{1}

\journalname{Nuclear Physics A}

\runauth{A. Beraudo et al.}

\jid{nupha}

\jnltitlelogo{Nuclear Physics A}

\usepackage{graphicx}
\usepackage{amsmath,amssymb}

\begin{document}

\begin{frontmatter}



\title{Heavy-flavour transport: from large to small systems}

\author{A. Beraudo, A. De Pace, M. Monteno, M. Nardi and F. Prino}

\address{INFN, Sezione di Torino, Via Pietro Giuria 1, I-10125 Torino}



\begin{abstract}
Predictions for heavy-flavour production in relativistic heavy-ion experiments provided by the POWLANG transport setup, including now also an in-medium hadronization model, are displayed, After showing some representative findings for the Au-Au and Pb-Pb cases,  a special focus will be devoted to the results obtained in the small systems formed in proton(deuteron)-nucleus collisions, where recent experimental data suggest the possible formation of a medium featuring a collective behaviour.
\end{abstract}


\end{frontmatter}



\section{Introduction}\label{sec:intro}
During the last few years we developed a tool to study heavy-flavour (HF) observables in high-energy hadronic collisions, in which one expects to produce a hot deconfined medium, able to modify the momentum and angular distribution not only of light hadrons, but also of heavy particles. In particular, we interfaced to a standard pQCD event generator (the POWHEG-BOX package~\cite{Alioli:2010xd}, supplemented, when necessary, by EPS09~\cite{Eskola:2009uj} nuclear corrections to the PDF's), used for the simulation of the initial $Q-\overline{Q}$ production, a transport code (developed by us) based on the relativistic Langevin equation, employed to describe the propagation of $c$ and $b$ quarks in the Quark-Gluon Plasma. The evolution of the latter was provided by relativistic hydrodynamic calculations~\cite{Romatschke:2007mq,DelZanna:2013eua}. Recently, we included in such a setup the possibility of accounting for medium modifications of heavy-quark hadronization. The procedure, based on the recombination of the heavy quarks with light thermal partons, leading to the formation of colour-singlet strings and their subsequent fragmentation, was described in detail in Ref.~\cite{Beraudo:2014boa}. Overall, it turned out to provide a better description of the experimental data, the collective motion inherited from the light quarks increasing the radial and elliptic flow of HF hadrons. Here, we wish to extend our previous studies, addressing also possible modifications of HF observables in small systems, like the ones produced in d-Au collisions at RHIC or in p-Pb collisions at the LHC. For more details on this last issue we refer the reader to Ref.~\cite{Beraudo:2015wsd}. 

\section{Heavy-flavour production in A-A collisions}\label{sec:AA}
\begin{figure}[!ht]
\begin{center}
\includegraphics*[clip,width=0.46\textwidth]{RAA_D0_RHIC_POW+string_nPDF_HTLvslQCD.eps}
\includegraphics*[clip,width=0.53\textwidth]{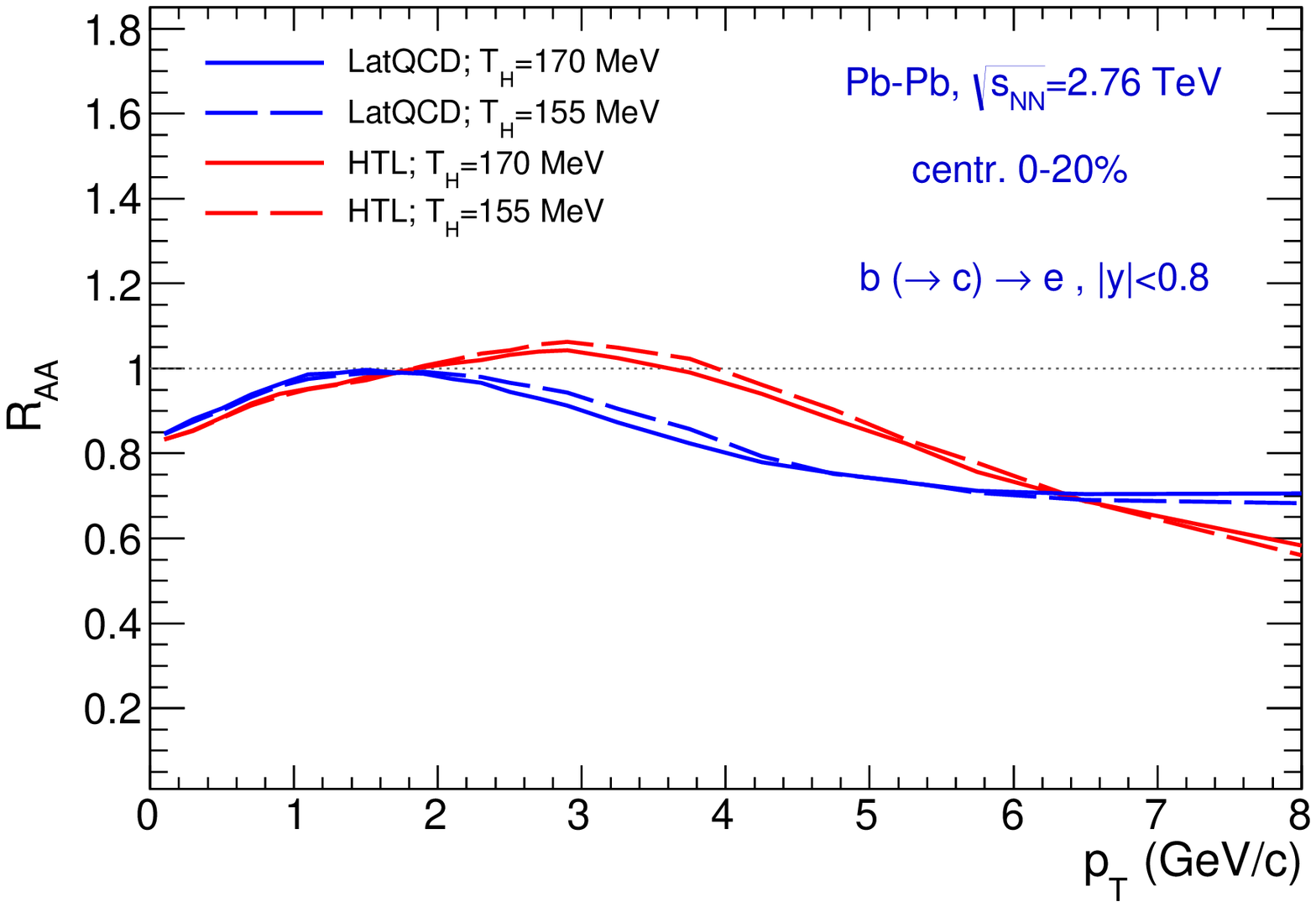}
\caption{POWLANG outcomes for the nuclear modification factor of D-mesons in Au-Au collisions at RHIC and of the electrons from beauty decays in Pb-Pb collisions at the LHC. Results with different transport coefficients and heavy-flavour decoupling temperatures are compared.}
\label{fig:AA}
\end{center}
\end{figure}
The improved version of our transport setup, including in-medium hadronization, was employed to study several HF observables in A-A collisions. A wide set of outcomes of our model (POWLANG) for single-particle momentum and angular distributions and two-particle correlations can be found in Ref.~\cite{Beraudo:2014boa}. 
As a representative example of the results we obtained, in the left panel of Fig.~\ref{fig:AA} we show POWLANG predictions for the nuclear modification factor ($R_{\rm AA}$) of D mesons in Au-Au collisions at $\sqrt{s_{\rm NN}}\!=\!200$ GeV at RHIC. Results obtained with different transport coefficients (from weak-coupling Hard-Thermal-Loop calculations~\cite{Alberico:2013bza} or lattice-QCD simulations~\cite{Francis:2015daa}) and HF decoupling temperatures are displayed and compared to STAR data~\cite{Adamczyk:2014uip}. The most evident feature in the data is the bump around $p_T\!\approx\!1.5$ GeV/c, which is qualitatively reproduced by our calculations for most of the scenarios explored in our analysis. Within our framework, D-meson spectra are affected both by the rescatterings of the heavy quarks crossing the QGP and by the additional collective flow acquired at hadronization from the light partons of the deconfined medium, whose importance has been stressed also in independent studies~\cite{He:2011qa,Cao:2013ita,Gossiaux:2009mk}: the bump in the D-meson $R_{\rm AA}$ is thus interpreted as a signature of the radial flow of the fireball.

In the right panel of Fig.~\ref{fig:AA} we move to Pb-Pb collisions at $\sqrt{s_{\rm NN}}\!=\!2.76$ TeV at the LHC and we display our new results for the the nuclear modification factor of electrons from beauty decays, for which preliminary ALICE results have recently become available~\cite{Volk:2015qoa}. As for the previous case, various transport coefficients and decoupling temperatures are explored.

\section{Heavy-flavour production in small systems}\label{sec:small}
One of the most surprising findings in the experimental search for the Quark-Gluon Plasma was certainly the signature of possible collective effects, suggestive of the formation of a hot strongly-interacting medium, recently observed in the collisions of small systems like p-Pb at the LHC and d-Au (and now also $^3$He-Au) at RHIC, in particular when selecting events characterized by a high multiplicity of produced particles. Various observables support the above picture: the structure of two-particle correlations in the $\Delta\eta\!-\!\Delta\phi$ plane (double ridge), suggestive of a boost-invariant initial condition, with azimuthal spatial asymmetries mapped by the strong interactions into the final particle spectra; the non-vanishing values of the elliptic, triangular and higher flow-harmonics, which seem to indicate a common correlation of all the particles with the same symmetry-plane; the hardening of the $p_T$-spectra moving towards more central events, which can be described as the effect of the collective radial flow of an expanding medium. The above effects display also a characteristic dependence on the particle species (mass ordering), still in agreement with the expectations of a hydrodynamic description.

\begin{figure}[!ht]
\begin{center}
\includegraphics*[clip,height=6cm]{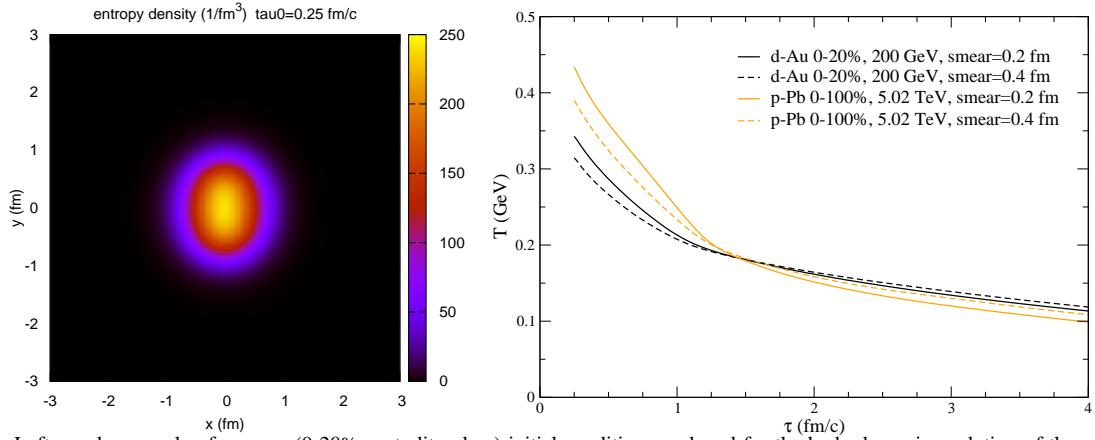}
\includegraphics*[clip,width=0.48\textwidth]{temperature.eps}
\caption{Left panel: example of average (0-20\% centrality class) initial condition employed for the hydrodynamic evolution of the medium in p-Pb collisions at $\sqrt{s_{\rm NN}}\!=\!5.02$ TeV. Right panel: the evolution of the temperature at the center of the fireball in d-Au and p-Pb collisions.}
\label{fig:init}
\end{center}
\end{figure}
Would the small deconfined system possibly produced in such collisions leave its signatures also on HF observables? This is the question we have recently tried to answer. Here we summarize our approach and our major findings, referring the reader to Ref.~\cite{Beraudo:2015wsd} for a more detailed discussion.
We modeled the initial state of the hydrodynamic evolution of the medium through a Glauber Monte-Carlo approach, generating several thousands initial conditions, ordering them according to the number of binary nucleon-nucleon collisions (used as a centrality estimator) and taking the average of all the ones (each one rotated in the transverse plane in order to share the same elliptic symmetry-plane) belonging to the centrality percentile considered. Each nucleon-nucleon collision was assumed to deposit some entropy in the transverse plane, smeared according to a Gaussian distribution. The subsequent evolution of the fireball was then evaluated through the ECHO-QGP code~\cite{DelZanna:2013eua}. As shown in Fig.~\ref{fig:init}, for the collisions for which HF data are available at RHIC and LHC, this was sufficient to get a medium which, starting from an initial temperature around 400 MeV, lived in the deconfined phase for a time up 2-3 fm/c.

\begin{figure}[!ht]
\begin{center}
\includegraphics*[clip,width=0.42\textwidth]{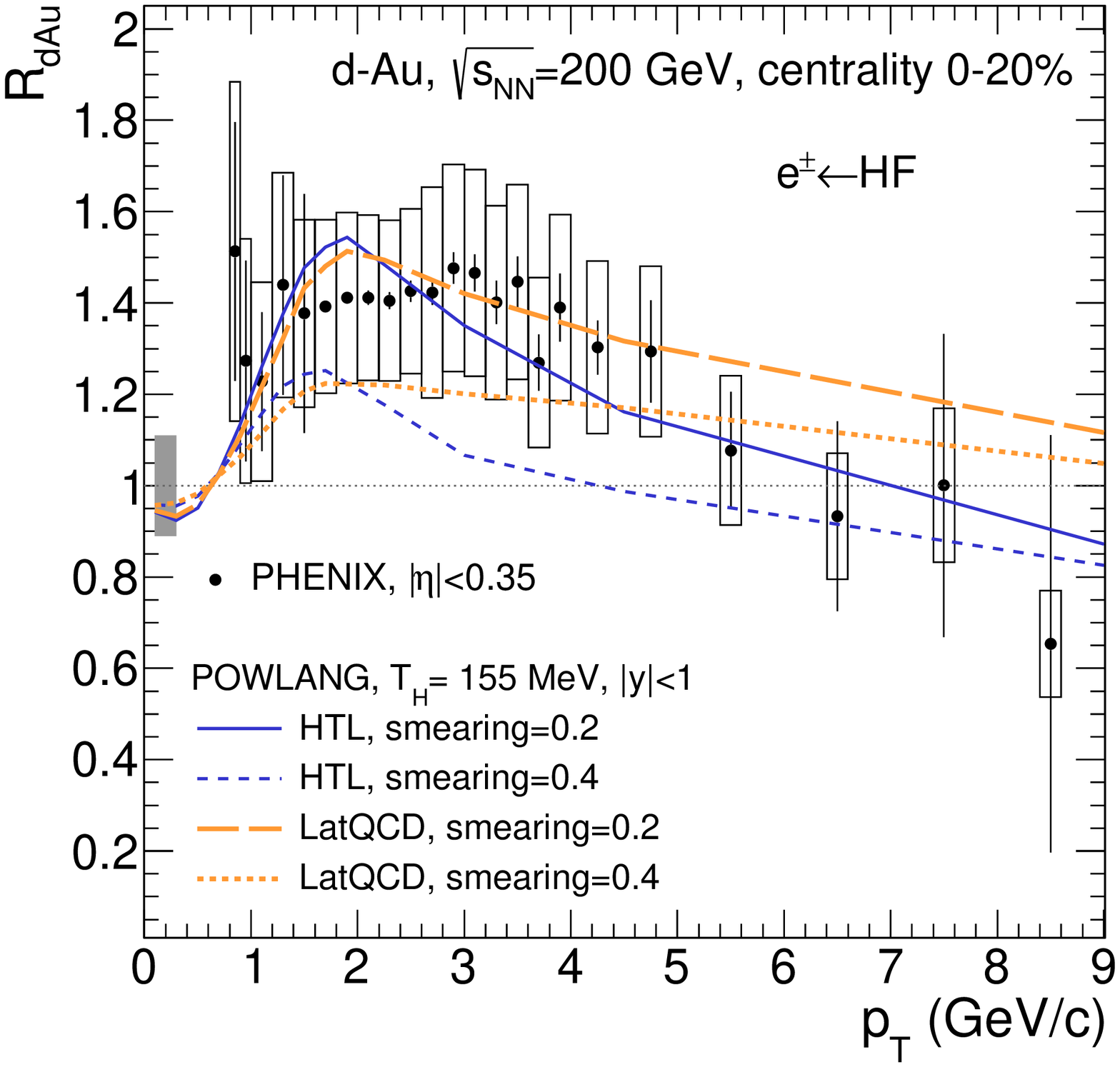}
\includegraphics*[clip,width=0.42\textwidth]{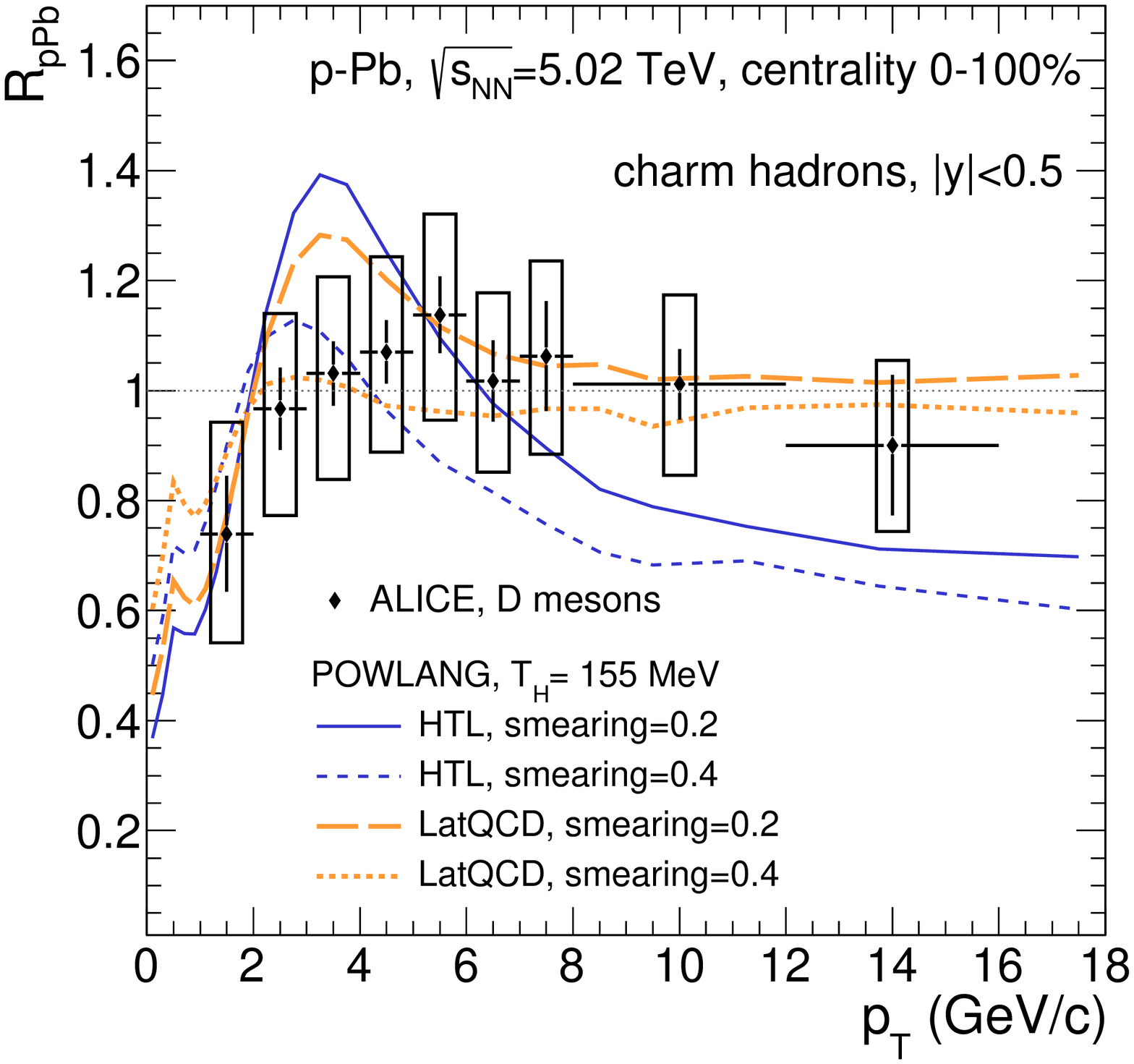}
\caption{POWLANG outcomes for the nuclear modification factor of HFE's in central d-Au collisions at RHIC and of D-mesons in p-Pb collisions at the LHC. Results with different initial conditions and transport coefficients are displayed and compared to PHENIX~\cite{Adare:2012yxa} and ALICE~\cite{Abelev:2014hha} data.}
\label{fig:small_thvsexp}
\end{center}
\end{figure}
As in the nucleus-nucleus case, the propagation and hadronization of the heavy quarks in the presence of such a deconfined medium was then simulated by our transport code. Here we display our main findings, obtained with different transport coefficients and initial conditions, for the currently accessible experimental observables. 
In the left panel of Fig.~\ref{fig:small_thvsexp} we show our predictions for the nuclear modification factor of electrons from heavy-flavour decays (HFE's) in central (0-20\%) d-Au collisions at $\sqrt{s_{\rm NN}}\!=\!200$ GeV. Our curves are compared to PHENIX results~\cite{Adare:2012yxa}. The main feature of the experimental data is that they overshoot unity over a quite extended $p_T$-range; within our setup this has to be interpreted -- as suggested also in Ref.~\cite{Sickles:2013yna} -- as due to the radial flow of the parent D and B mesons, partly developed at the partonic level, partly acquired at hadronization.  
In the right panel  of Fig.~\ref{fig:small_thvsexp} we focus on the $p_T$-spectra of charmed hadrons produced in p-Pb collisions at $\sqrt{s_{\rm NN}}\!=\!5.02$ TeV. POWLANG curves are compared to ALICE data for the average $R_{\rm pPb}$ of D mesons. The rescatterings of the heavy quarks in the medium and their recombination with light thermal partons at hadronization induce a depletion of the spectra at very low momentum, giving rise to a bump in the theory curves around $p_T\!\approx\!3$ GeV/c. More quantitative details of the results depend on the theoretical scenario considered. So far, experimental D-meson data do not allow one to draw firm conclusions about the existence of final-state hot-medium effects, although they do not look in contradiction with our results.    

\begin{figure}[!ht]
\begin{center}
\includegraphics*[clip,width=0.42\textwidth]{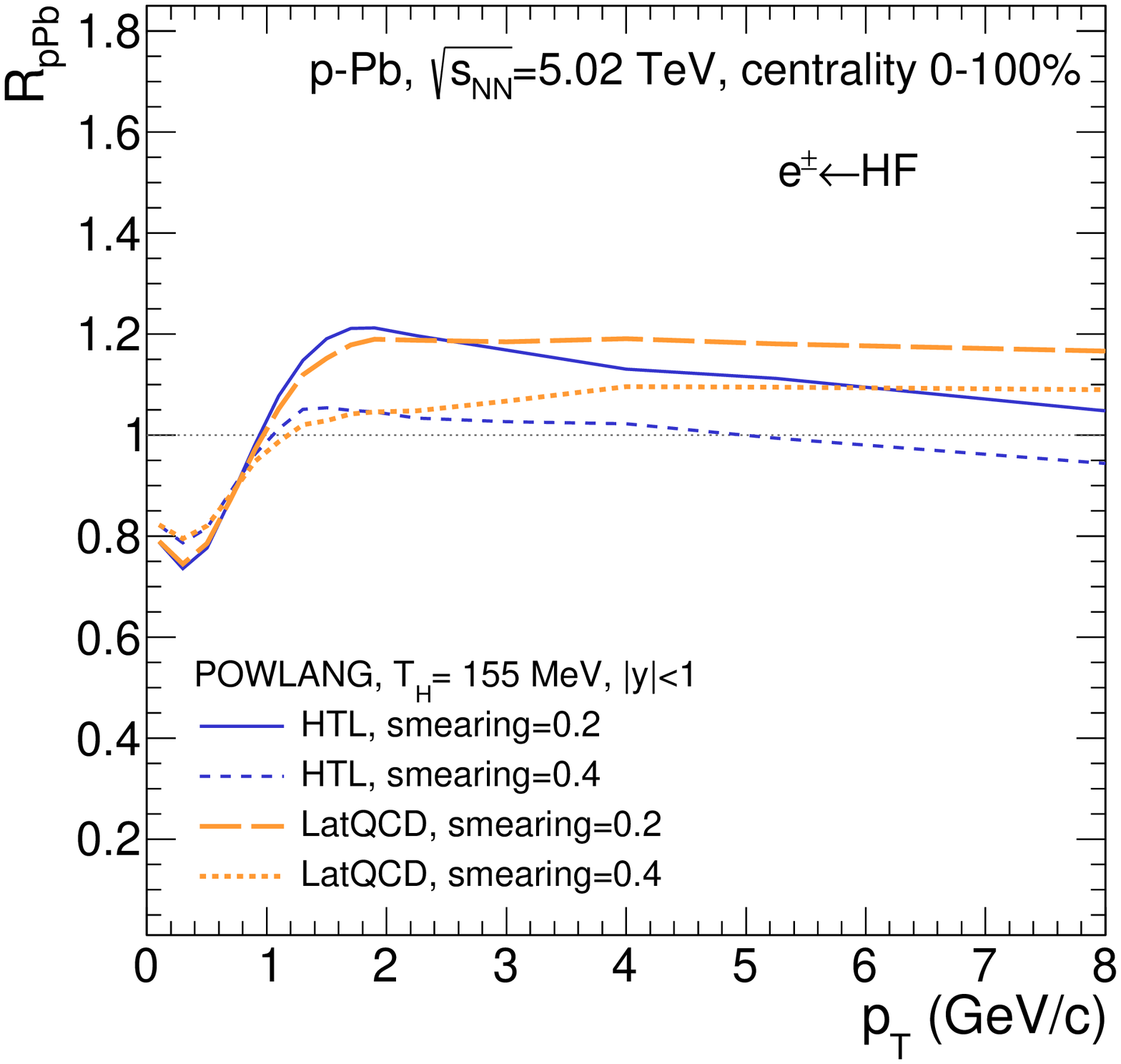}
\includegraphics*[clip,width=0.42\textwidth]{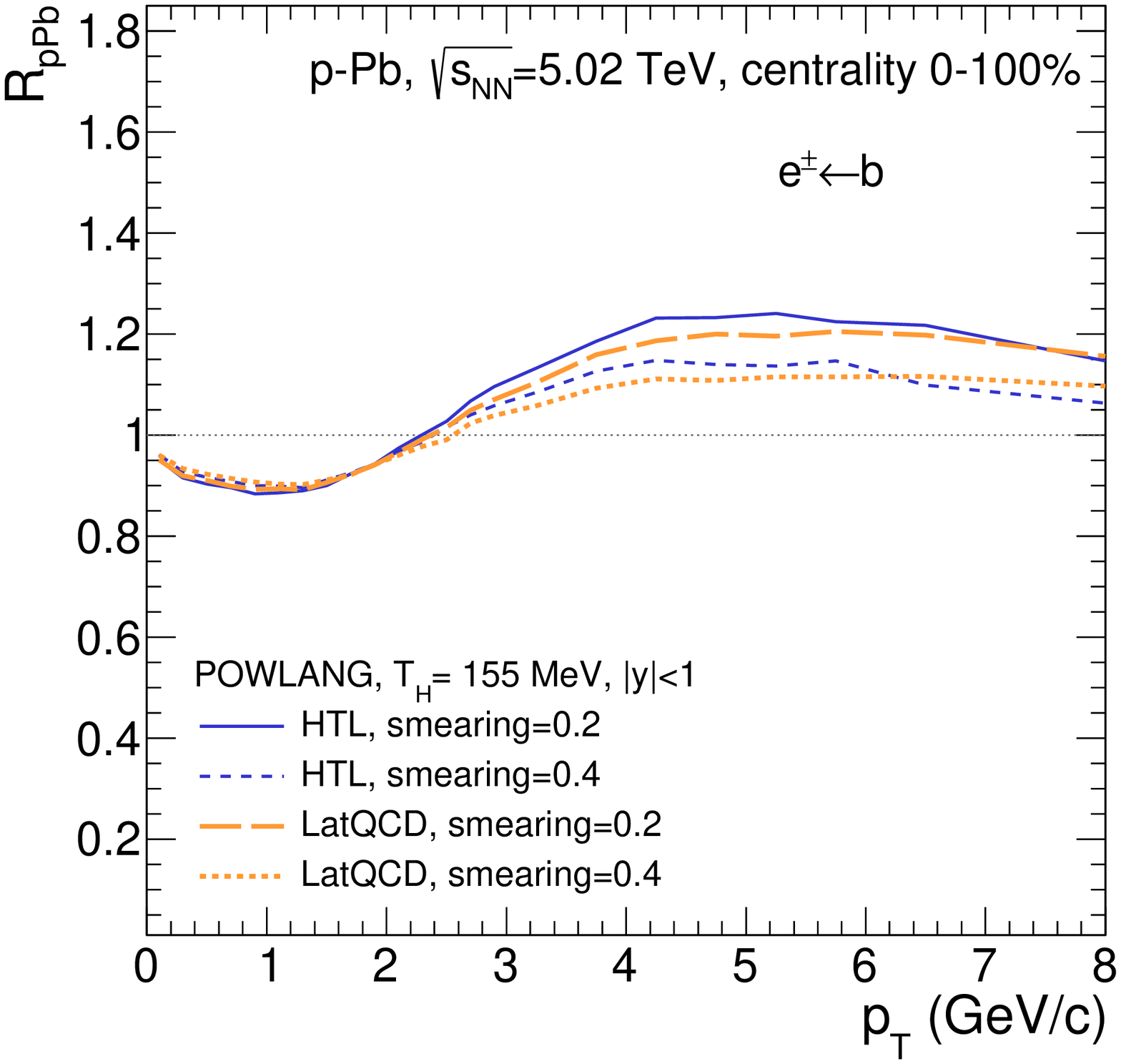}
\caption{POWLANG outcomes for the HFE $R_{\rm pPb}$ ($e_{c+b}$ in the left panel and $e_b$ in the right one) in p-Pb collisions at the LHC. Results with different initial conditions and transport coefficients are displayed.}
\label{fig:electrons}
\end{center}
\end{figure}
Finally, in Fig.~\ref{fig:electrons} we display POWLANG results for the nuclear modification factor of HFE's in p-Pb collisions at $\sqrt{s_{\rm NN}}\!=\!5.02$ TeV, recently studied by the ALICE collaboration~\cite{Adam:2015qda}. Both the inclusive results and the separate beauty contribution are plotted. Overall, model calculations display a depletion at small momenta and tend to slightly overshoot unity for moderate $p_T$. 

\bibliography{paper}

\begin{thebibliography}{10}
\expandafter\ifx\csname url\endcsname\relax
  \def\url#1{\texttt{#1}}\fi
\expandafter\ifx\csname urlprefix\endcsname\relax\def\urlprefix{URL }\fi
\expandafter\ifx\csname href\endcsname\relax
  \def\href#1#2{#2} \def\path#1{#1}\fi

\bibitem{Alioli:2010xd}
S.~Alioli, P.~Nason, C.~Oleari, E.~Re, JHEP 06 (2010) 043.
\newblock \href {http://arxiv.org/abs/1002.2581} {\path{arXiv:1002.2581}}.

\bibitem{Eskola:2009uj}
K.~J. Eskola, H.~Paukkunen, C.~A. Salgado, JHEP 04 (2009) 065.
\newblock \href {http://arxiv.org/abs/0902.4154} {\path{arXiv:0902.4154}}.

\bibitem{Romatschke:2007mq}
P.~Romatschke, U.~Romatschke, Phys. Rev. Lett. 99 (2007) 172301.
\newblock \href {http://arxiv.org/abs/0706.1522} {\path{arXiv:0706.1522}}.

\bibitem{DelZanna:2013eua}
L.~Del~Zanna, G.~others, A.~Drago, F.~Becattini, Eur. Phys. J. C73 (2013) 2524.
\newblock \href {http://arxiv.org/abs/1305.7052} {\path{arXiv:1305.7052}}.

\bibitem{Beraudo:2014boa}
A.~Beraudo, A.~De~Pace, M.~Monteno, M.~Nardi, F.~Prino, Eur. Phys. J. C75~(3)
  (2015) 121.
\newblock \href {http://arxiv.org/abs/1410.6082} {\path{arXiv:1410.6082}}.

\bibitem{Beraudo:2015wsd}
A.~Beraudo, et~al.\href {http://arxiv.org/abs/1512.05186}
  {\path{arXiv:1512.05186}}.

\bibitem{Alberico:2013bza}
W.~M. Alberico, et~al., Eur. Phys. J. C73 (2013) 2481.
\newblock \href {http://arxiv.org/abs/1305.7421} {\path{arXiv:1305.7421}}.

\bibitem{Francis:2015daa}
A.~Francis, et~al., Phys. Rev. D92~(11) (2015) 116003.
\newblock \href {http://arxiv.org/abs/1508.04543} {\path{arXiv:1508.04543}}.

\bibitem{Adamczyk:2014uip}
L.~Adamczyk, et~al., Phys. Rev. Lett. 113~(14) (2014) 142301.
\newblock \href {http://arxiv.org/abs/1404.6185} {\path{arXiv:1404.6185}}.

\bibitem{He:2011qa}
M.~He, R.~J. Fries, R.~Rapp, Phys. Rev. C86 (2012) 014903.
\newblock \href {http://arxiv.org/abs/1106.6006} {\path{arXiv:1106.6006}}.

\bibitem{Cao:2013ita}
S.~Cao, G.-Y. Qin, S.~A. Bass, Phys. Rev. C88 (2013) 044907.
\newblock \href {http://arxiv.org/abs/1308.0617} {\path{arXiv:1308.0617}}.

\bibitem{Gossiaux:2009mk}
P.~B. Gossiaux, R.~Bierkandt, J.~Aichelin, Phys. Rev. C79 (2009) 044906.
\newblock \href {http://arxiv.org/abs/0901.0946} {\path{arXiv:0901.0946}}.

\bibitem{Volk:2015qoa}
M.~Völk, J. Phys. Conf. Ser. 612~(1) (2015) 012037.

\bibitem{Adare:2012yxa}
A.~Adare, et~al., Phys. Rev. Lett. 109~(24) (2012) 242301.
\newblock \href {http://arxiv.org/abs/1208.1293} {\path{arXiv:1208.1293}}.

\bibitem{Abelev:2014hha}
B.~B. Abelev, et~al., Phys. Rev. Lett. 113~(23) (2014) 232301.
\newblock \href {http://arxiv.org/abs/1405.3452} {\path{arXiv:1405.3452}}.

\bibitem{Sickles:2013yna}
A.~M. Sickles, Phys. Lett. B731 (2014) 51--56.
\newblock \href {http://arxiv.org/abs/1309.6924} {\path{arXiv:1309.6924}}.

\bibitem{Adam:2015qda}
J.~Adam, et~al.\href {http://arxiv.org/abs/1509.07491}
  {\path{arXiv:1509.07491}}.

\end{thebibliography}

\end{document}